# Frictional scattering and frictional waveguides: achieving persistent superlubricity at high velocity on the nanoscale

Yilun Liu[1], François Grey[1,2,3,*] and Quanshui Zheng[1,&]

[1]*Department of Engineering Mechanics, Tsinghua University, Beijing 100084, P. R. China*

[2]*Department of Physics, Tsinghua University, Beijing 100084, P. R. China*

[3] *London Centre for Nanotechnology, University College London, London WC1H OAH, U.K.*

*e-mail: grey@mail.tsinghua.edu.cn   &e-mail: zhengqs@tsinghua.edu.cn

**Nanomechanical devices can operate at much higher speeds than their macroscopic analogues, due to low inertia. For example, peak speeds >100m/s have been predicted for carbon nanotube devices[1-4]. This stimulates our interest in the atomic-scale physics of friction at high velocity. Here we study a model nanosystem consisting of a graphene flake moving freely on a graphite substrate at >100m/s. Using molecular dynamics we discover that ultra-low friction, or superlubricity, is punctuated by high-friction transients as the flake rotates through successive crystallographic alignments with the substrate. We term this phenomenon *frictional scattering* and show that it is mathematically analogous to Bragg scattering. We also show that frictional scattering can be eliminated by using graphitic nanoribbons as *frictional waveguides* to constrain the flake rotation, thus achieving persistent superlubricity. Finally, we propose an experimental method to study nanoscale high-velocity friction. These results may guide the design of efficient high-frequency nanomechanical devices.**

The nonlinear nature of friction at the nanoscale has sparked intense experimental and theoretical interest[5-7]. A focus of much of this research is on achieving superlubricity[8-11], a term originally introduced to describe ultra-low friction between rotationally misaligned – and hence incommensurate – crystalline lattices[12,13]. Superlubricity has been observed by Friction Force Microscopy, which typically operates at <<1mm/s. However, for freely-moving nanoparticles at low velocities, superlubricity is suppressed because of rotation into energetically favourable commensurate states[15-16]. Energetic considerations suggest that superlubricity should be favoured at sufficiently high velocities, when kinetic energy overcomes the periodic substrate potential[11]. Simulations for one-dimensional systems confirm this, while revealing a rich variety of resonant frictional phenomena[17].

To investigate a two-dimensional nanoscale graphene flake at velocities >100m/s, we employ molecular dynamics simulations using the Gromacs code[18] with the Drieding force field[19]. A graphite substrate is modeled by two layers of graphene in A-B stacking with the bottom layer fixed, and periodic boundary conditions applied to a 30nm×20nm region. Results presented in Fig. 1 are for a 10x10nm$^2$ flake launched at $v_{max}$=400m/s along the armchair direction of the substrate. Flake and substrate have an initial T=0K configuration and A-B stacking alignment. The simulations are in time steps of 1fs, corresponding to <0.0004nm at $v_{max}$. The total energy of flake and substrate is constant, and we ignore electronic contributions to friction[20].

Fig. 1b shows the trajectory of the flake, with sporadic changes in the direction of motion indicated by vertical dashed lines, corresponding to frictional scattering. Snapshots in Fig. 1a show each scattering occurs when the flake is crystallographically aligned with the substrate. Fig. 1c shows the component of the velocity along the x-direction, $v_x$, dropping in discrete steps at each scattering, with similar stepwise variations for $v_y$. The evolution of flake rotation angle, plotted in Fig. 1d, indicates that the flake angular velocity, $\omega$, also changes abruptly at each scattering. Fig. 1e shows that scattering events subject the flake to large oscillatory forces in both x and y directions, producing abrupt changes of direction and velocity. In Fig. 1f, the linear kinetic energy components $E_x$ and $E_y$ in the $x$- and $y$-directions and the rotational kinetic energy $E_\omega$ of the flake are displayed. Since $E_y$ and $E_\omega$ remain small throughout, we conclude that the initial kinetic energy of the flake is primarily dissipated into vibrational energy, with enhanced dissipation at each scattering. After ~1ns, the flake is effectively thermalized, with $E_x$ and $E_y$ reaching similar magnitudes, and the flake moving randomly.

In Fig. 2a, we show data for five flakes with sizes between 4x4nm$^2$ and 10x10nm$^2$. The overall behaviour is similar, all flakes being thermalized within 2ns. However, the duration of intervals of superlubricity, $\tau_s$, varies widely and seems uncorrelated with flake size. To understand this, in Fig. 2b we plot $\tau_s$ against $\omega$ during the corresponding interval. The data follow a simple relationship $\tau_s = 60°/\omega$. In other words, the rate of frictional scattering is simply determined by the rate of crystallographic alignments.

Since the total energy of the system remains constant, we estimate the temperature after thermalization to be <30K. This heating effect is exaggerated by the boundary conditions. We have studied the sensitivity of the results to initial temperature and flake orientation, see Supplementary Figure 1. We have also tested other flake shapes and varied substrate boundary conditions. In all cases we observe a similar stepwise decay, indicating that this is a computationally robust effect.

To study the effect more closely, we zoom in on a region of frictional scattering in Fig. 3. The flake velocity component, $v_x$, and van der Waals bonding energy between flake and substrate, $E_{vdW}$, are displayed in Figs. 3a-b, respectively, and show rapid oscillations near alignment. Fig. 3c shows eight snapshots of the moiré pattern generated by the overlap of flake and substrate lattices, at different points indicated in Figs. 3a-b. When the flake is rotationally misaligned with the substrate, the moiré pattern has a small unit cell size, $a_m$. This means the spatial phase of flake atoms relative to the substrate potential oscillates rapidly across the flake. The result is a very small net corrugation potential for the flake as a whole: this is the origin of superlubricity for incommensurate lattices[21].

As the flake rotates into alignment, $a_m$ expands (snapshots 1-4). Near alignment, once $a_m \gg l$, where $l$ is the linear size of the flake, all flake atoms are nearly in phase with the substrate (snapshots 5,6) and the net corrugation potential experienced by the flake increases abruptly, resulting in scattering. The corrugation potential drops again once the flake rotates sufficiently far out of alignment (snapshots 7,8). The variation of the corrugation potential, $E_c$, can clearly be seen in Figs. 3d-g. $E_c$ is estimated along the *x*-direction before, during and after alignment, by displacing the flake relative to its instantaneous position, and recalculating $E_{vdW}$ keeping all other parameters fixed. Near alignment (snapshots 5, 6) $E_c$ is ~100x larger than during superlubricity, and the flake is subject to large potential gradients as indicated in Figs. 3e,f.

In the reference frame of the moving flake, the fundamental Fourier component of the substrate potential along the x-direction appears as a travelling wave of wavelength $\lambda=d$, where *d* is the in-plane lattice constant of graphite. When rows of atoms in the flake align with crests of this wave, there is a maximum oscillatory force acting on the flake. This corresponds to the condition for Bragg scattering $2d\sin\theta=n\lambda$, at normal incidence with *n*=2. (For odd *n*, the force on neighbour rows cancels out.)

Since frictional scattering depends on the crystallinity of the flake, it will not be observed for diffusion of single atoms or small molecules at similar velocities[22]. Also, since frictional scattering depends on rotation in a plane, it is an intrinsically two-dimensional effect, physically distinct from velocity-dependent resonances observed in one-dimensional models[17] and nanotube simulations[23]. This may explain why, to our knowledge, the phenomenon has not been reported before.

We can derive the angular width $\Delta\theta_c$ of the coherent region where $a_m \gg l$ using the standard moiré fringe formula[24] $a_m = d/(2|\sin(\Delta\theta/2)|)$, where $\Delta\theta$ is the rotation angle relative to alignment. Substituting for $a_m = 2l$ as a lower bound for coherence across the flake gives $\Delta\theta = d/2l$ in the small angle limit. Since $\Delta\theta$ is symmetric around the aligned position, $\Delta\theta_c = d/l$. This is identical to the Scherrer equation for a crystallite of dimension $l$ and radiation of wavelength $\lambda = d$, for the case of normal incidence[25], emphasizing the mathematical analogy with Bragg scattering. Substituting $d=0.246$nm, the lattice constant of graphite and $l=10$nm gives $\Delta\theta_c = 1.4°$. This angular width is plotted in Fig. 3b, and agrees with the range where $E_{vdW}$ oscillates rapidly.

Based on the above analysis, we propose suppressing frictional scattering by constraining the rotational motion of the flake. As shown in Fig. 4a, we launch a 10x10nm$^2$ flake on a graphitic nanoribbon of width 11nm, with the flake rotated 90° relative to the normal AB stacking of the nanoribbon. Fig. 4b shows that after 2ns, the velocity of the flake decays by less than 1%, compared with complete decay of the forward motion for the unconstrained flake. The nanoribbon acts as a frictional waveguide, periodically reflecting the flake off its edges, producing velocity oscillations shown in the inset, and limiting flake rotation to ±1°. However, the waveguide has a velocity cutoff: smooth sliding ceases below ~15m/s, a typical velocity for transition to stick-slip motion in atomic-scale simulations[26].

To probe high-velocity friction experimentally, a mechanism is required to accelerate a graphene flake to >100m/s. Inspired by the ability of scanning probe microscopes (SPMs) to manipulate nanoparticles[27] and the observation of self-retraction of microscopic graphite flakes after shearing[28], we propose pushing a nanoscale flake of graphitic material partially over a step edge on a graphite surface and releasing it. In Fig. 5 a 10x10nm$^2$ graphene flake, initially extending a distance $s$=5nm over a step edge, rapidly accelerates to ~600m/s after release. This acceleration is due to reduction of free surface energy, as also predicted for telescoping nanotubes[1,2]. From classical mechanics, it is straightforward to deduce an expression for the peak velocity produced by this effect: $v_{max} = [(2/N)\,(s/l)\,(E_x/m_C)]^{1/2}$ where $N$ is the number of graphitic layers in the flake, $E_x$ is the exfoliation energy per carbon atom and $m_C$ is the mass of a carbon atom. Setting $s/l$=1/2, $N$=1 and $E_x$ = 53meV[29] gives $v_{max}$ = 658m/s, in good agreement with simulation. Although SPM cannot capture such high-speed motion directly, the resulting displacement of the flake should be readily detectable.

In conclusion, the motion of a nanoscale graphene flake on graphite at high velocity is dominated by frictional scattering, where large oscillatory forces act each time the flake rotates through crystallographic alignment with the substrate. Frictional scattering is intrinsically two dimensional, and can be suppressed by constraining the flake to move along a quasi-one-dimensional frictional waveguide. These insights should stimulate further experimental and theoretical interest in high-velocity friction, a regime of considerable practical importance for future nanomechanical devices.

References

1. Cumings, J. & Zettl, A. Low-friction nanoscale linear bearing realized from multiwall carbon nanotubes. *Science* **289**, 602-604 (2000).

2. Zheng, Q.S. et al. Self-retracting motion of graphite microflakes. *Phys. Rev. Lett.* **100** 067205 (2008).

3. Barreiro, A. et al. Subnanometer motion of cargoes driven by thermal gradients along carbon nanotubes. *Science* **320**, 775-778 (2008).

4. Somada, H., Hirahara, K., Akita, S. & Nakayama, Y. A molecular linear motor consisting of carbon nanotubes. *Nano Lett.* **9**, 62-65 (2009).

5. Urbakh, M., Klafter, J., Gourdon, D. & Israelachvili J. The nonlinear nature of friction, *Nature* **430**, 525-528 (2004)

6. Luan, B., & Robbins, M.O. The breakdown of continuum models of mechanical contacts *Nature* **435**, 929-932 (2005)

7. Mo, Y., Turner, K.T. & Szlufarska, I. Friction laws at the nanoscale *Nature* **457**, 1116-1119 (2009)

8. Socoliuc, A., Bennewitz, R., Gnecco & Meyer, E. Transition From Stick-Slip to Continuous Sliding in Atomic Friction: Entering a New Regime of Ultralow Friction *Phys. Rev. Lett.* **92**, 134301 (2004)

9. Socoliuc, A. et al. Atomic-scale control of friction by actuation of nanometer-sized contacts. *Science* **313**, 207-210 (2006).

10. Lantz, M.A., Wiesmann, D. & Gotsmann, B. Dynamic superlubricity and the elimination of wear on the nanoscale, *Nature Nanotech.* **4**, 586-591 (2009)

11. Hirano, M., Superlubricity of Clean Surfaces. *Superlubricity* (eds. Erdemir A. & Martin J.M., Elsevier 2007)

12. Hirano, M., Shinjo, K., Kaneko & R., Murata, Y. Anisotropy of Frictional Forces in Muscovite. *Phys. Rev. Lett.* **67**, 2642-2645 (1991).

13. Shinjo, K. & Hirano M. Dynamics of friction: superlubric state. *Surf. Sci.* **283**, 473-478 (1993).

14. Dienwiebel, M. et al. Superlubricity of graphite. *Phys. Rev. Lett.* **92** 126101 (2004).

15. Depondt, Ph., Ghazali, A. & Levy, J.-C.S. Self-locking of modulated single overlayer in a nanotribology experiment. *Surf. Sci.* **419**, 29-37 (1998).

16. Filippov, A.E., Dienwiebel, M., Frenken, J.W.M., Klafter, J. & Urbakh, M. Torque and twist against superlubricity. *Phys. Rev. Lett.* **100** 046102 (2008).

17. Weiss, M. & Elmer F.-J. Dry friction in the Frenkel-Kontorova-Tomilinson model: dynamical properties. *Z. Phys. B* **104**, 55-69 (1997)

18. Lindahl, E., Hess, B. & van der Spoel, D. GROMACS 3.0: a package for molecular simulation and trajectory analysis. *Journal of Molecular Modeling* **7**, 306-317 (2001).

19. Guo, Y.J., Karasawa, N. & Goddard, W.A. Prediction of Fullerene Packing in C60 and C70 Crystals. *Nature* **351**, 464-467 (1991).

20. Krim, J., Daly, C. & Dayo, A. Electronic contributions to sliding friction *Trib. Lett.* **1**, 211-218 (1995).

21, Sacco, J.E. & Sokoloff, J.B. Free sliding in solids with two incommensurate periodicities. *Phys. Rev. B* **18**, 6549 (1978)

22. Hedgeland, H. et al. Measurement of single-molecule frictional dissipation in a prototypical nanoscale system *Nature Phys.* **5** 561-564 (2009)

23. XU, Z.P. et al. Trans-phonon effects in ultra-fast nanodevices *Nanotechnology* **19** 255705-255709 (2008)

24. Woan, G. The Cambridge *Handbook of Physics Formulas* (Cambridge Uni. Press 2000)

25. Dinnebier, R.E. & Billinge, S.J.L. *Powder Diffraction: Theory and Practice* (Royal Society of Chemistry, 2008).

26. Braun, O.M. et al. Transition from smooth sliding to stick-slip motion on a single frictional contact. *Phys. Rev. E* **72** 056116 (2005)

27. Dietzel, D. et al. Frictional duality observed during flake sliding. *Phys. Rev. Lett.* **101** 125505 (2008).

28. Zheng, Q.S. et al. Self-retracting motion of graphite microflakes. *Phys. Rev. Lett.* **100** 067205 (2008).

29. Girifalco, L.A. & Lad, R.A. Energy of Cohesion, Compressibility, and the Potential Energy Functions of the Graphite System. *Journal of Chemical Physics* **25**, 693-697 (1956).

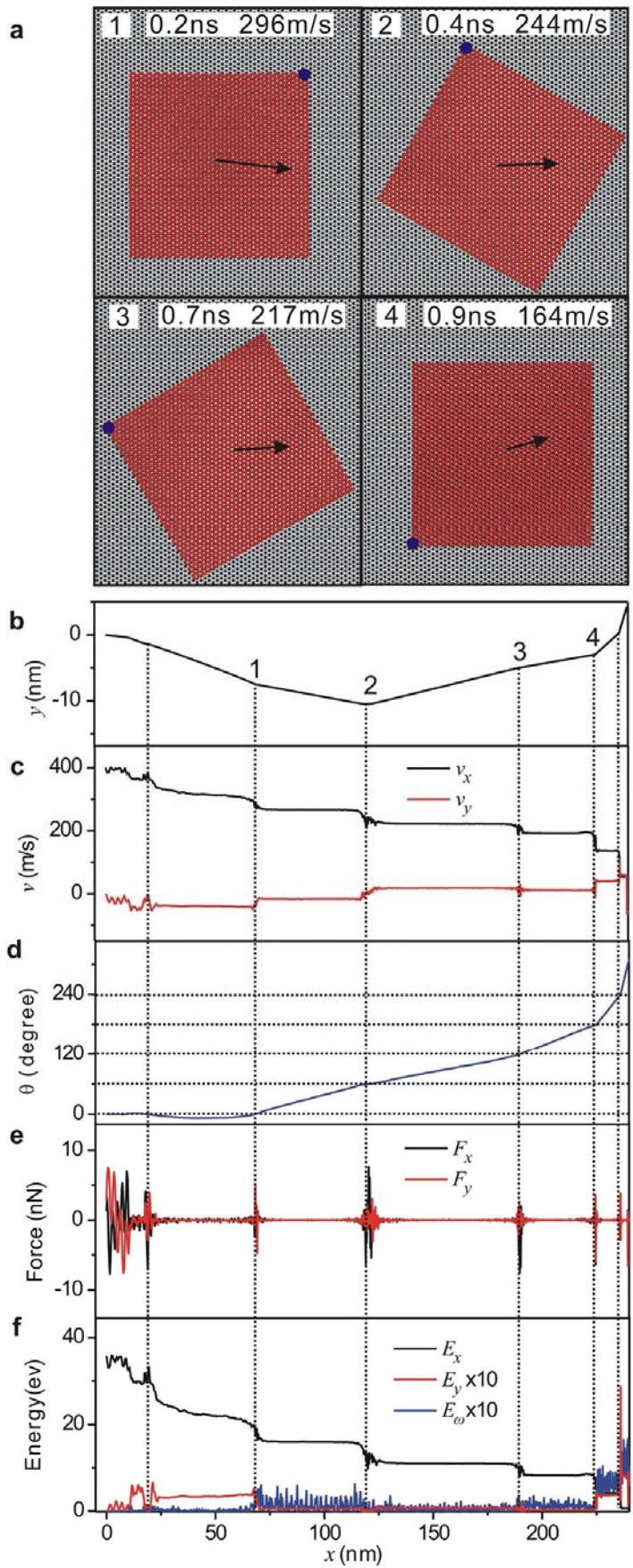

**Figure 1 | Flake dynamics after launch at 400m/s.** Four snapshots (**a**) of the

instantaneous orientation and velocity of the flake during frictional scattering at four points indicated on the flake trajectory (**b**). The velocity components of the flake $v_x$ and $v_y$ (**c**), show stepwise changes at each scattering. The flake rotation $\theta$ (**d**) shows frictional scattering occurring primarily at high-symmetry angles (multiples of 60°). The force components acting on the flake, $F_x$ and $F_y$ (**e**) show large oscillatory variations at each scattering. The kinetic energy component $E_x$ decays within about 1ns to less than 10% of its initial value (**f**), and becomes comparable to $E_y$ and $E_\omega$. The flake begins to execute random displacements at this point.

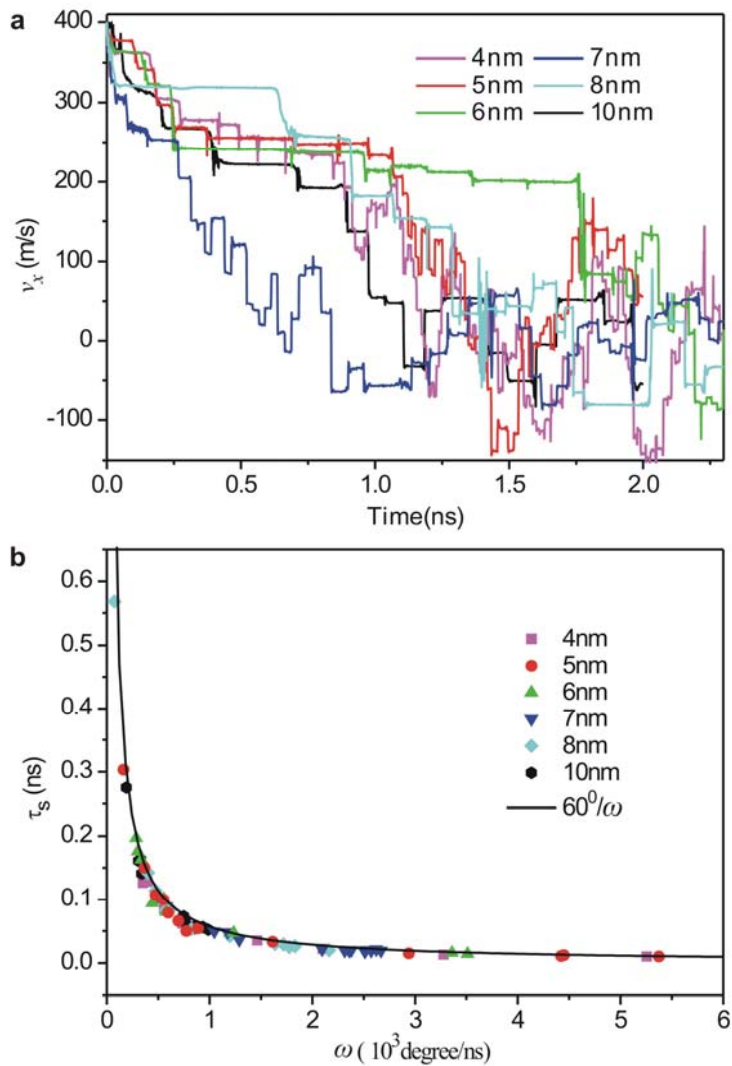

**Figure 2 | Dependence of persistent superlubricity on angular velocity.** Comparison of decay of the velocity component $v_x$ versus time after launch for different flake sizes (**a**) shows similar behavior for all sizes, although with significant variations in the duration of intervals of persistent superlubricity. This is explained by the correlation (**b**) between the duration of intervals of superlubricity, $\tau_s$, and the average angular velocity of the flakes during the intervals, ω, showing that $\tau_s$ depends inversely on the angular velocity.

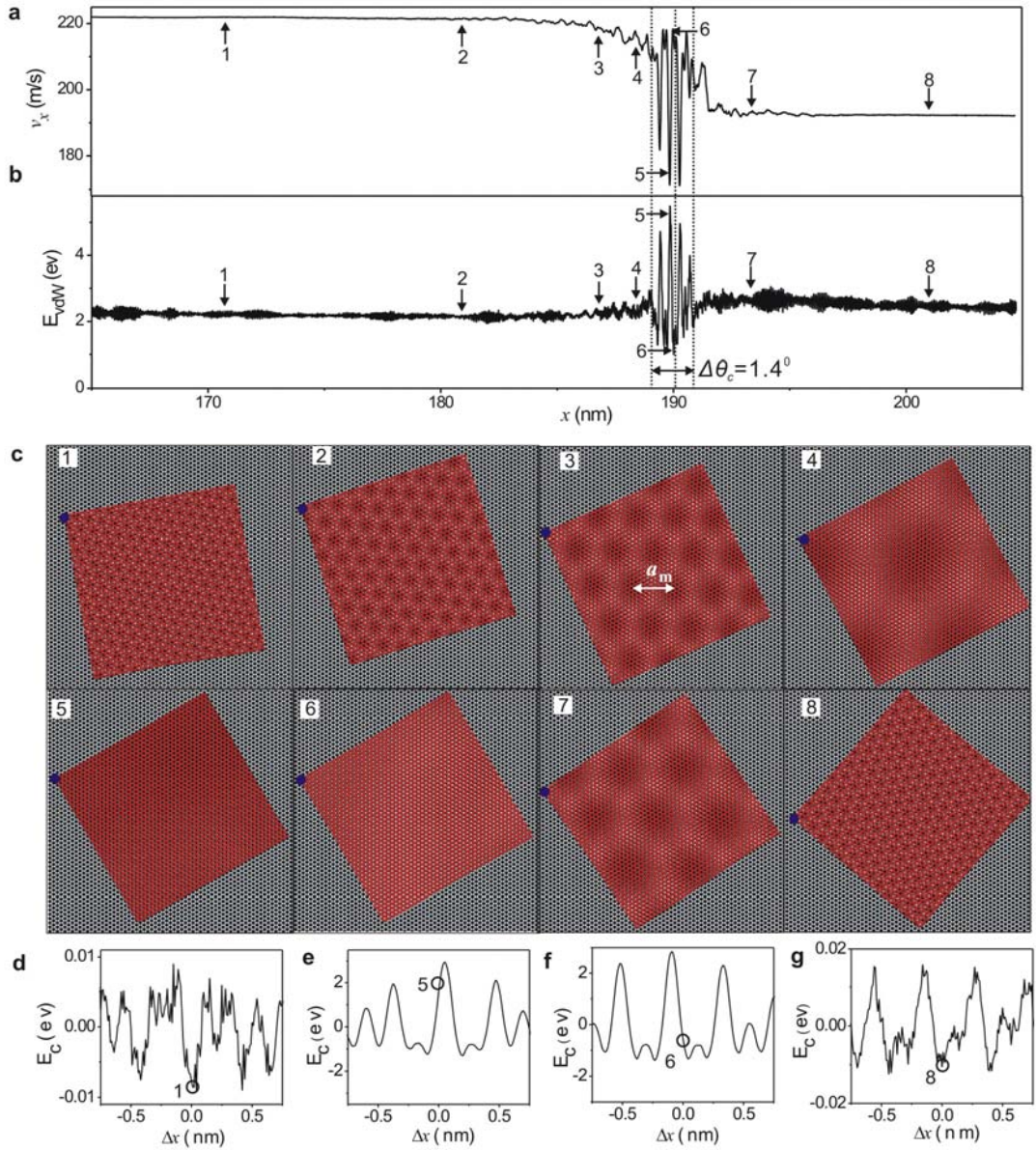

**Figure 3 | Close-up of frictional scattering.** Variation of the velocity component $v_x$ (**a**) and the van der Waals bonding energy $E_{vdW}$ (**b**) as a function of flake displacement along the x-direction, before, during and after a scattering event. Numbers indicate the snapshots of the flake (**c**), which track the variation of the moiré pattern generated by the overlapping periodicities near crystalline alignment between flake and substrate. For snapshots 5 and 6 the flake atoms are nearly in phase with the

substrate, and the moiré pattern is no longer visible because its unit cell $a_m$, indicated in snapshot 3, becomes much larger than the linear dimension of the flake, $l$. Vertical dotted lines in (**a**) and (**b**) indicate the angular width $\Delta\theta_c$ of the coherent region where $a_m > 2l$, the central dotted line corresponding to crystallographic alignment of flake and substrate. The instantaneous corrugation potential, $E_c$, felt by the flake, for relative displacement in the *x*-direction about the flake's actual position (indicated by a circle) is shown before (**d**), during (**e,f**) and after (**g**) frictional scattering. During scattering, the corrugation potentials is ~100x greater than during superlubricity, and the flake experiences very large forces, as evidenced by the large potential gradients in (**e,f**).

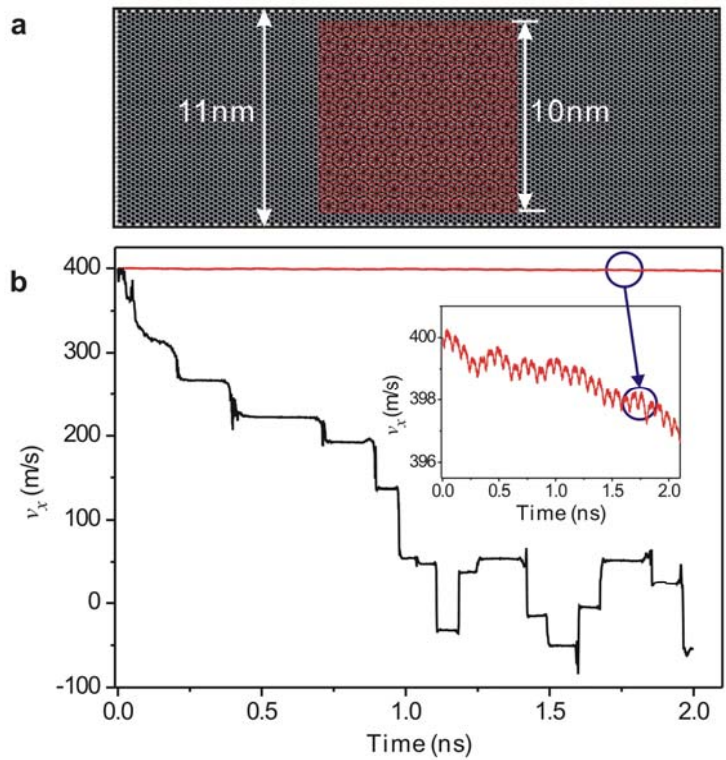

**Figure 4 | Suppression of frictional scattering using a graphitic nanoribbon.** After release at 400m/s, the snapshot (**a**) shows how a 10x10nm$^2$ flake initially rotated 90° relative to the substrate is rotationally constrained as it moves along a graphitic nanoribbon 11nm wide. As a result, in a plot of velocity component $v_x$ versus time (**b**), no frictional scattering is observed (upper red curve), in contrast to the rapid decay of velocity for an unconstrained flake (lower black curve). The inset shows the same data with the velocity scale greatly magnified for the red curve, revealing a velocity decay of less than 1% in 2ns. A small oscillation of the forward velocity visible in the inset corresponds to the flake reflecting off the two edges of the nanoribbon as it moves forward.

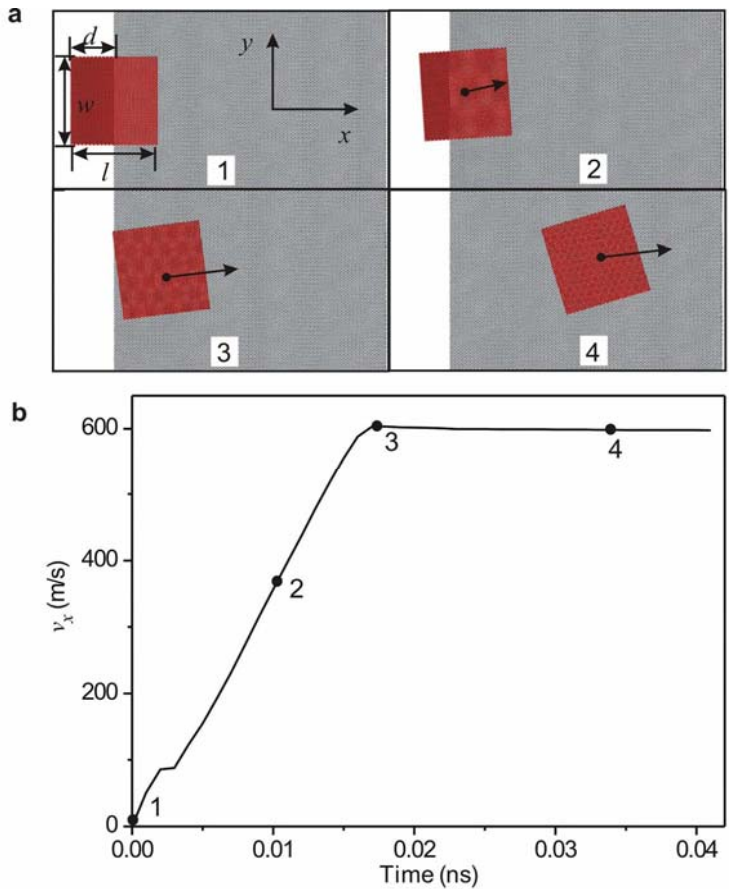

**Figure 5 | Acceleration of a graphene flake on a graphite substrate.** Four snapshots (**a**) of a 10x10nm2 flake initially extending 5nm over the edge of a graphite substrate, and a plot of the velocity component $v_x$ vs. time after release of the flake (**b**), showing rapid acceleration to approximately 600m/s. Although the flake starts in the aligned high-friction state it almost immediately rotates, so that superlubricity predominates during the acceleration.

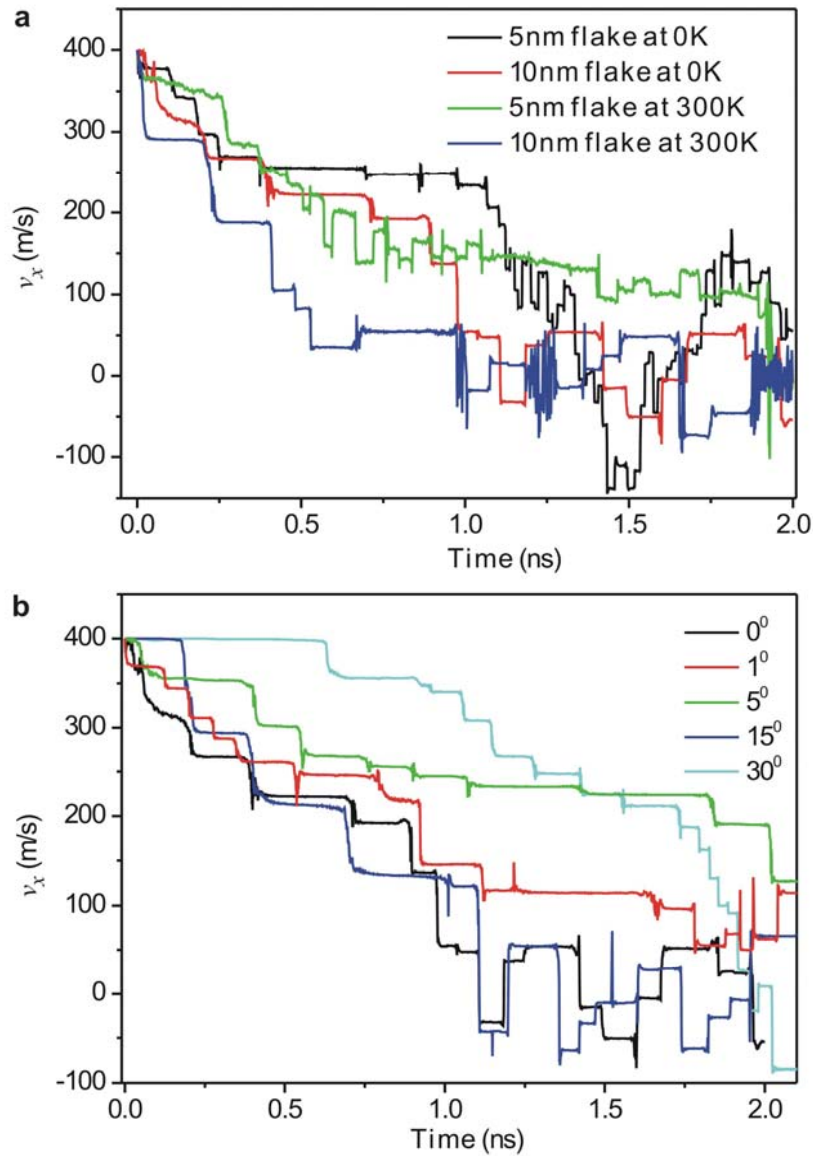

**Supplementary Figure 1 | Dependence of results on temperature and initial orientation of the flake.** By introducing an initial displacement pattern in both flake and substrate atoms, corresponding to vibrational excitiations at finite temperature, it is possible to test the robustness of persistent superlubricity to temperature change (**a**). The result for 5x5nm$^2$ and 10x10nm$^2$ flakes launched initially at 0K and 300K show similar qualitative behavior, with regions of superlubricity punctuated by frictional scattering. Detailed inspection shows a higher degree of noise

for the high-temperature simulations, which is consistent with expectations for the impact of temperature. Simulations at T=0K, but for different initial orientations of the flake when it is launched, show overall similar qualitative behavior, with the velocity of all flakes decaying within 2ns. The flakes rotated furthest from the low-energy alignment orientation (15° and 30°) spend longer in an initial superlubricity state, which is also consistent with expectations, since they must rotate a larger angle before frictional scattering occurs.